\documentclass[12pt]{iopart}
\usepackage{graphicx}
\usepackage{iopams}
\newcommand{\W}{\mathcal{W}}

\newcommand{\QQ}{\mathcal{Q}}
\newcommand{\RR}{\mathcal{R}}
\newcommand{\hatrho}{\hat{\rho}}
\newcommand{\hatgamma}{\hat{\gamma}}
\newcommand{\hatH}{\hat{H}}
\newcommand{\hatsigma}{\hat{\sigma}}
\newcommand{\llangle}{\langle\!\langle}
\newcommand{\rrangle}{\rangle\!\rangle}
\newcommand{\rlangle}{\rangle\!\langle}
\newcommand{\rrllangle}{\rrangle\!\llangle}
\begin{document}

\title[The influence of charge detection on counting statistics]{The influence of charge detection on counting statistics}{$5$-th International Conference on
Unsolved Problems on Noise}
\author{Alessandro Braggio}
\address{LAMIA-INFM-CNR,
             Dipartimento di Fisica,
             Universit\`{a} di Genova,
             Via Dodecaneso 33, 16146 Genova, Italy}
\ead{braggio@fisica.unige.it}
\author{Christian Flindt}
\address{Department of Physics, Harvard University, 17 Oxford Street, Cambridge, MA 02138, USA}
\address{Department of Condensed Matter Physics,
             Faculty of Mathematics and Physics, Charles University,
             Ke Karlovu 5, 12116 Prague, Czech Republic}
\author{Tom\'{a}\v{s} Novotn\'{y}}
\address{Department of Condensed Matter Physics,
             Faculty of Mathematics and Physics, Charles University,
             Ke Karlovu 5, 12116 Prague, Czech Republic}
             
\begin{abstract}
We consider the counting statistics of electron transport through a
double quantum dot with special emphasis on the dephasing induced by a nearby charge detector. The double dot is embedded in a dissipative enviroment, and the presence of electrons on the double dot is detected with a nearby quantum point contact. Charge transport through the double dot is governed by a non-Markovian generalized master equation. We describe how the cumulants of the current can be obtained for such problems, and investigate the difference
between the dephasing mechanisms induced by the quantum point contact and the coupling to the external heat bath. Finally, we consider various open questions of relevance to future research.
\end{abstract}

\maketitle

\section{Introduction}
The study of random fluctuations has a relevant role in many branches
of physics~\cite{Vanderziel:1959,Kogan:1996,Wax2003,Nazarov:2003}. Close to equilibrium, fluctuations are intimately connected with dissipative relaxation mechanisms according to the fluctuation-dissipation theorem, independently of the
physical origin of the fluctuations, classical or quantum-mechanical~\cite{Montroll:1979,Vankampen:2007}.
In contrast, far from equilibrium fluctuating quantities provide a unique insight into the internal properties of the
system under consideration~\cite{Zwanzig:2001}. An immediate example is the evaluation of the
quasi-particle charge of carriers through measurements of the current cumulants~\cite{Braggio:2006b}.

Many important phenomena can be characterized in terms of counted, elementary
entities. The concept of particles, in the quantum realm, naturally defines what
quantities should be counted. From this point of view one recognizes that
counting
problems constitute a quite general framework in which many different dynamical processes can be interpreted,
also in the
presence of complex quantum physics. The first application of the counting approach
in quantum
physics came from photon counting experiments, where the concept of full
counting
statistics (FCS) was originally developed~\cite{Mandel:1995}. Recently, this concept
has attracted intensive
theoretical~\cite{Nazarov:2003} and
experimental~\cite{Reulet:2003,Bomze:2005,Fujisawa:2006}
attention within the
field of electron transport. In the context of mesoscopic transport, FCS was
introduced in order to characterize the noise properties of nanodevices~\cite{Levitov:1996}. Later, it was
demonstrated also to
be a sensitive diagnostic tool for detecting quantum-mechanical coherence,
entanglement,
disorder, and dissipation~\cite{Nazarov:2003}.

Mathematically, FCS encodes the complete knowledge of the probability distribution
$P(n,t)$ of the number
$n$ of transmitted entities during the measurement time $t$ or, equivalently, of all
corresponding cumulants. The study of counting statistics for stochastic processes is generally of
broad relevance for a wide class of problems. For example, non-zero higher order
cumulants describe
non-Gaussian behavior and contain information about rare events, whose study has become
an
important topic within non-equilibrium statistics in physics, chemistry, and biology \cite{Vankampen:2007,Bardou:2002,Shapiro:2003}.
Within the framework of master equations some important results were recently obtained. Bagrets and Nazarov~\cite{Bagrets:2003}
have shown that the cumulant
generating function (CGF) corresponding to a Markovian master equation
is determined by
the dominating eigenvalue of the rate matrix, when counting fields are appropriately
included. Some of us have shown that it in principle is
possible to
calculate arbitrary orders of cumulants using perturbation theory in
the counting
field, rather than solving  the full eigenvalue problem for the dominating eigenvalue \cite{Flindt:2005}. For non-Markovian systems
described
by generalized master equations (GME),
we have
shown that
the CGF scales linearly with time  \cite{Braggio:2006,Flindt:2008}, as in the case of Markovian processes, if the memory kernel has no power-law tails. Moreover, the CGF can be calculated using a
so-called
non-Markovian expansion \cite{Braggio:2006}.

Recently, we developed a method which unifies and
extends
these earlier approaches to FCS within a
GME formulation \cite{Flindt:2008}. Due to their
intrinsic analytic structures, the previous approaches were in practice limited to systems with only a
few states \cite{Bagrets:2003,Braggio:2006}, or only the first few current
cumulants could be addressed \cite{Flindt:2005}.  In contrast, this recent advancement enables
studies of a much larger class of problems, including the evaluation of
zero-frequency current cumulants of very high orders for non-Markovian systems
with many states~\cite{Flindt:2008}.
We also showed how the method allows calculations of the finite-frequency
current noise for non-Markovian transport processes \cite{Flindt:2008,Aguado:2004a}. A detailed account of these techniques will be given elsewhere \cite{Flindt:2008a}.
In this paper we mainly want to restate the essential findings and address the open questions of relevance in future research.

In order to demonstrate the applicability of our methods, we consider the current
fluctuations of charge transport through a coherently coupled double quantum dot, a charge qubit. We consider the effects of a nearby
quantum point contact (QPC) charge detector and the coupling to a dissipative phonon bath.
If the QPC barrier height is modulated by electrons on the double quantum dot,
the fluctuations of the
current through
the QPC monitors the dynamics of the qubit. This introduces a
qubit dephasing mechanism. As we shall see,
current fluctuations can be useful for
extracting  information about the internal dynamics of the double dot
system. We
concentrate on the transition between coherent and incoherent transport through the qubit,
demonstrating the sensitivity of the cumulants to this transition.

The structure of the paper is as follows: In Section \ref{sec:NMGME} we summarize the general concepts of our method, while clearly identifying the essential steps to obtain the
FCS or the current cumulants for a system governed by a non-Markovian GME. We
briefly sketch the derivation of the expressions for the first few current cumulants (current, noise, and skewness)
used in the following section.
In Section \ref{sec:Model} we describe the model of a dissipative qubit with a nearby QPC charge detector and show
how the current cumulants yield information
about the dynamics of the qubit. In Section \ref{sec:Open} we discuss various open questions and give an outlook for future research.

\section{Non-Markovian GME}
\label{sec:NMGME}

For many nanoscopic systems it is convenient to consider the evolution of only a few degrees of freedom.
The system evolution is then captured by the dynamic equation for the reduced density
matrix corresponding to these degrees of freedom. This equation should contain the effects of all external forces driving the
system and, at the same time, the effective dynamics due to the degrees of freedom that have been traced
out. The dynamics of the reduced system is, in general, non-Markovian, and can for a large class of processes be described by a generic non-Markovian GME of the form
\cite{Zwanzig:2001,Flindt:2008,Makhlin:2001}
\begin{equation}
\frac{d}{dt}\hatrho(n,t)=\sum_{n'}\int_{0}^{t}dt'\W(n-n',t-t')\hatrho(n',t')+
\hatgamma(n,t).
\label{eq:GME}
\end{equation}
Here, the reduced density matrix $\hatrho(n,t)$ of the system has been resolved with respect to the number of transferred charges $n$. The memory kernel $\W$ describes the
influence of the environment on the dynamics of the system, while the inhomogeneity
$\hatgamma$ accounts for initial correlations between system and
environment.\footnote{With environment we refer to all degrees of freedom
that have been traced out.} For a given system, the derivation of an equation like Eq.\ (\ref{eq:GME}) may be a
difficult task, but many different examples can already be found in the literature \cite{Bagrets:2003,Plenio:1998,Flindt:2007}.
Both $\W$ and $\hatgamma$ decay with time, usually on a comparable timescale.
We consider systems where $\W$
and $\hatgamma$ with time decay faster than any power-law.
With this condition it can be shown that the effects of the inhomogeneity vanish in the long-time limit.
The inhomogeneity $\hatgamma$ is consequently irrelevant for all statistical quantities in the long-time limit. For finite times, it may, however, play
a crucial role \cite{Flindt:2008}.

The cumulant
generating function $\mathcal{S}(\chi,t)$ corresponding to $P(n,t)$ is
defined as
\begin{equation}
\label{eq:CGF}
e^{\mathcal{S}(\chi,t)}=\sum_{n}P(n,t)e^{in\chi}=\sum_{n} \mathrm{Tr}\{\hatrho(n,t)\}
e^{in\chi}
\end{equation}
where $\chi$ is the so-called counting field. The second equality
defines $P(n,t)$ as the trace of the $n$-resolved reduced density matrix.
The $m$'th cumulant $\llangle
n^m\rrangle(t)$ is directly connected to the Taylor coefficients of the CGF
in Eq.\ (\ref{eq:CGF}) according to the definition
$\llangle n^m\rrangle(t)\equiv\partial^m \mathcal{S}(\chi,t)/\partial(i \chi)^m|_{\chi\rightarrow 0}$.
We now derive a general expression for the CGF of a system described by a GME of the form given in Eq.\
(\ref{eq:GME}). In Laplace space, defined by the transform
$f(\chi,z)\equiv\sum_{n}\int_0^{\infty}dt f(n,t)e^{in\chi-zt}$,
the equation has the algebraic form
\begin{equation}
z\hatrho(\chi,z)-\hatrho(\chi,t=0)=\W(\chi,z)\hatrho(\chi,z)+\hatgamma(\chi,z),
\end{equation}
which can formally be solved for $\hatrho(\chi,z)$ by introducing the
resolvent $\mathcal{G}(\chi,z)\equiv[z-\W(\chi,z)]^{-1}$. Returning to the time domain by an inverse Laplace transformation,
 the CGF  becomes
\begin{equation}
e^{\mathcal{S}(\chi,t)}=\frac{1}{2\pi i}\int_{a-i\infty}^{a+i\infty}dz
\mathrm{Tr}\left\{\mathcal{G}(\chi,z)[\hatrho(\chi,t=0)+\hatgamma(\chi,z)]\right\}
e^{zt},
\label{eq:GF}
\end{equation}
where $a$ is a real number, chosen such that all singularities of
the integrand are situated to the left of the vertical line of
integration. This expression constitutes a powerful
formal result, but as we shall see in the following, it also leads to useful practical schemes.

As already mentioned above, the CGF scales linearly with time in the long-time limit for kernels that decay
faster than any power-law and is independent of the initial conditions \cite{Braggio:2006}. For such systems we can define
the zero-frequency cumulants of the current as $\llangle
I^m\rrangle=\frac{d}{dt}\llangle
n^m\rrangle(t)|_{t\rightarrow\infty}$ (with $e=1$ in the following).
With the counting field $\chi$ set to zero, the system tends exponentially to a unique
stationary state determined by the $1/z$ pole of the resolvent
$\mathcal{G}(\chi=0,z)$. The stationary state is given by the
eigenvector corresponding to the zero-eigenvalue of
$\W_0\equiv\W(\chi=0,z=0)$, i.e., $\lim_{t\rightarrow
\infty}\hatrho(\chi=0,t)\equiv|0\rrangle$, where $|0\rrangle$ is
the normalized solution to $\W_0|0\rrangle=0$. With finite
values of $\chi$, an eigenvalue $\lambda_0(\chi,z)$ develops
adiabatically from the zero-eigenvalue and the long-time behavior is still determined by the
pole $1/[z-\lambda_0(\chi,z)]$ of $\mathcal{G}(\chi,z)$ close to
zero. The particular pole $z_0(\chi)$ that solves the self-consistency equation
\begin{equation}
z_0-\lambda_0(\chi,z_0)=0,
\label{eq:self-consistency}
\end{equation}
and goes to zero with $\chi$ going to zero, i.e., $z_0(0)=0$,
consequently determines the long-time limit of the CGF. We thus find
$e^{\mathcal{S}(\chi,t)}\rightarrow \mathcal{D}(\chi)e^{z_0(\chi)t}$
for large $t$, where $\mathcal{D}(\chi)$ is a time-independent
function depending on the initial conditions. The current cumulants
then read $\llangle I^m\rrangle=\frac{\partial^m
z_0(\chi)}{\partial{(i\chi)^m}}|_{\chi\rightarrow 0}$. In the
Markovian limit for the kernel  $\W(\chi,z\rightarrow 0)$ we obtain
$z_0(\chi)=\lambda_0(\chi,0)$, consistently with previous results for the Markovian case \cite{Bagrets:2003,Flindt:2005}.
From Eq.\ (\ref{eq:self-consistency}), it is clear that calculations of current cumulants proceed in two steps: First,
 the dominating eigenvalue
$\lambda_0(\chi,z)$ has to be determined. Secondly, the self-consistency equation must be solved for $z_0(\chi)$.

In the original approach to calculations of CGF for non-Markovian transport systems a
so-called non-Markovian
expansion was developed \cite{Braggio:2006}. In this approach, the
CGF is expressed as a series using only the Taylor expansion of the dominating
eigenvalue around $z=0$.
Mathematically, the non-Markovian expansion is equivalent to the
solution of the self-consistency equation. The method, however,
requires an
analytic solution for the dominating eigenvalue with its full dependence on the counting field \cite{Braggio:2006}.
This is not a feasible approach, when the involved matrices are
large or, equivalently, for systems with many degrees of freedom.
However, if we are only interested in a finite number of cumulants an alternative route exists. To this end, we
have developed a scheme for calculating finite orders of current cumulants, where both
the eigenvalue problem and the
self-consistency equation are solved using perturbation theory in the counting field. A particular strength of the scheme
is that it is recursive, allowing for calculations of cumulants of very high orders \cite{Flindt:2008,Flindt:2008a}.
Here, we do not present all
details of the derivation of the recursive scheme, but
mainly focus on the final results for the first three current cumulants; mean current,
noise, and skewness.

We consider an expansion of the dominating eigenvalue in $\chi$ and $z$, $\lambda_0(\chi,z)=\sum_{k,l=0}^{\infty}\frac{(i\chi)^k}{k!}\frac{z^l}{l!}c^{(k,l)}$.
Let us first suppose that the coefficients $c^{(k,l)}$ are known. This allows us to solve Eq.\ (\ref{eq:self-consistency}) for $z_0(\chi)$ to a given
order in $\chi$. From the expansion
$z_0(\chi)=\sum_{n=1}^{\infty}\frac{(i\chi)^n}{n!}\llangle
I^n\rrangle$, we then extract the zero-frequency cumulants of the current. The results for the
lowest cumulants are
\begin{eqnarray}
\label{eq:I}
\llangle
I^1\rrangle =& c^{(1,0)},\\
\label{eq:II}
\llangle I^2\rrangle =& c^{(2,0)}+2c^{(1,0)}c^{(1,1)},\\
\llangle I^3\rrangle =& c^{(3,0)}+3 c^{(2,0)}c^{(1,1)}+3c^{(1, 0)}\left[c^{(1, 0)} c^{(1, 2)}+2(c^{(1,1)})^2+ c^{(2,1)}\right].
\label{eq:III}
\end{eqnarray}
Higher-order cumulants are readily calculated in a recursive manner \cite{Flindt:2008,Flindt:2008a}. The cumulants consist of contributions from
the purely Markovian quantities $c^{(k,0)}$ and the
non-Markovian terms $c^{(k,l)}$ for $0<l$. We observe the general rule that the $n$'th cumulant requires knowledge of non-Markovian terms $c^{(k,l)}$
of order $0<l<n$ \cite{Braggio:2006}. Consequently, the mean current is not
sensitive to non-Markovian effects, whereas higher-order cumulants
are. In general, the distinction made here between Markovian and
non-Markovian systems is related to the existence of a finite memory in Eq.\ (\ref{eq:GME}).
In practice, the memory effects depend on the choice of degrees of freedom that are traced out. It may for example be convenient
to trace out degrees of freedom of a Markovian system and describe the resulting, reduced system by
a non-Markovian GME as it was done in Ref.\ \cite{Flindt:2007}. Of course, in such cases our method yields identical results regardless of the particular formulation, Markovian or non-Markovian.

We still need to calculate the expansion coefficients
$c^{(k,l)}$ entering the expression $\lambda_0(\chi,z)=\sum_{k,l=0}^{\infty}\frac{(i\chi)^k}{k!}\frac{z^l}{l!}c^{(k,l)}$. The eigenvalue problem for $\lambda_0(\chi,z)$ is defined by the equation
\begin{equation}
\label{eq:eigen}
\W(\chi,z)|0(\chi,z)\rrangle=[\W_0+\W'(\chi,z)]|0(\chi,z)\rrangle=
\lambda_0(\chi,z)|0(\chi,z)\rrangle,
\end{equation}
where we have written $\W(\chi,z)$  as the sum of an unperturbed part $\W_0$ and the perturbation $\W'(\chi,z)\equiv\W(\chi,z)-\W_0$.
The right eigenvector $|0(\chi,z)\rrangle$ reduces adiabatically to the
stationary state $|0\rrangle$ with $\chi$ and $z$  going to zero and satisfies the normalization
condition $\llangle \tilde{0} |0(\chi,z)\rrangle=1$. Here, the left
eigenvector $\llangle \tilde{0}|$ is
defined by the relation $\llangle \tilde{0}|\W_0=0$. Using Rayleigh-Schr\"{o}dinger
perturbation theory we can express
the coefficients in terms of the Taylor coefficients of $\W'(\chi,z)$ denoted as $\W^{(k,l)}$, i.e.,
 $\W'(\chi,z)=\sum_{k,l=0}^{\infty}\frac{(i\chi)^k}{k!}\frac{z^l}{l!}\W^{(k,l)}$ with $\W^{(0,0)}=0$ by definition,
and the pseudoinverse of the kernel $\mathcal{R}$ defined as $\mathcal{R}\equiv\QQ\W_0^{-1}\QQ$. Even if $\W_0$ is singular, the pseudoinverse is in fact
well-defined since the singular part of the kernel is projected away by the projector $\QQ\equiv
1-|0\rrllangle\tilde{0}|$.\footnote{Details of the
super-operator notation used here can be found in Ref.~\cite{Flindt:2004}.}
We can now calculate the expansion coefficients and we report here the resulting expressions for a few of them
\begin{eqnarray}
& c^{(1,0)}=\llangle\tilde{0}|\W^{(1,0)}|0\rrangle,\\
& c^{(1,1)}=\llangle\tilde{0}|(\W^{(1,1)}-\W^{(1,0)}\RR\W^{(0,1)})|0\rrangle,\\
& c^{(2,0)}=\llangle\tilde{0}|(\W^{(2,0)}-2\W^{(1,0)}\RR\W^{(1,0)})|0\rrangle.
\end{eqnarray}
The expansion coefficients can also be calculated in a recursive manner as described in Refs.\ \cite{Flindt:2008,Flindt:2008a}.
We note that the recursive scheme can be used both for analytic and
numerical calculations. Evaluation of the pseudoinverse $\RR$ amounts to solving matrix
equations which is feasible even with very large matrices
\cite{Flindt:2004}. Numerically, the scheme is stable for
very high orders of cumulants ($>20$) as we have tested on simple
examples.

\section{Dissipative double quantum dot with a QPC charge detector}
\label{sec:Model}
\begin{center}
\begin{figure}
\includegraphics[width=.98\linewidth]{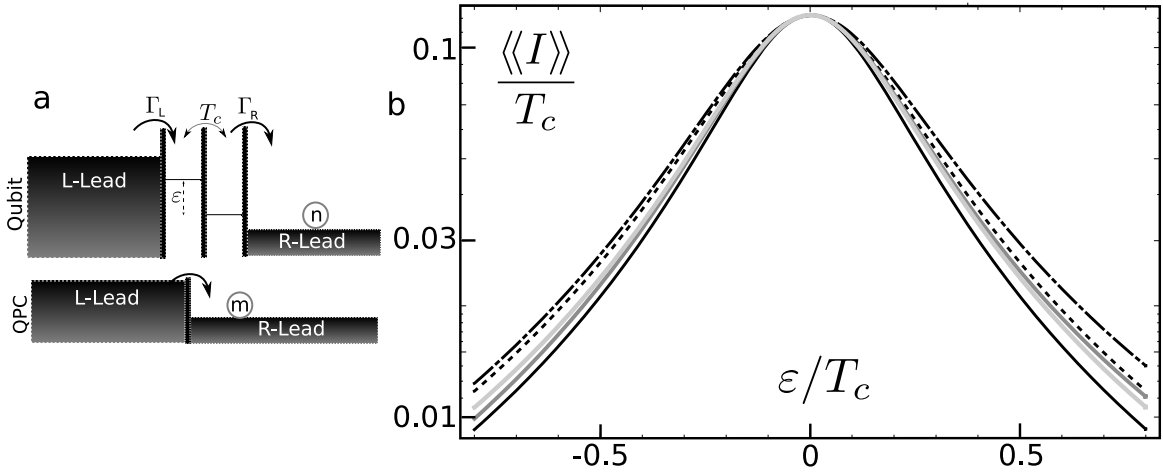}
\caption{a) Schematics of the double dot and the QPC charge detector. The state of the charge qubit
 modulates the transparency of the quantum point contact via the direct Coulomb interaction between them. b) Mean current
$\llangle I \rrangle/ T_c$ on a log-scale vs level detuning $\varepsilon/T_c$; line styles
correspond to different qubit regimes: coherent case ($\Gamma_d=\Gamma_R/2$, $\alpha=0$, solid),
coupled only to the QPC ($\alpha=0$, $\Gamma_d/T_c=0.075$, dotted), coupled only to the phonon-bath ($\Gamma_d=\Gamma_R/2$, $\alpha=6.25\times 10^{-4}$)
 for decoupling i) (dark grey) and decoupling ii)
(light grey), coupled to both QPC and
phonon bath for decoupling ii) (dot-dashed). Other parameters are
$\Gamma_L/T_c=10$, $\Gamma_L/T_c=10$ and, for gray and dot-dashed curves only, $k_B
T/T_c=5$. See text for descriptions of decouplings i) and ii).}
\label{fig:Model}
\end{figure}
\end{center}

We illustrate our method by considering a
model of charge transport through a double quantum
dot embedded in a dissipative environment, see Fig.~\ref{fig:Model}a. A QPC close to the double quantum dot is used as a charge
detector \cite{Gurvitz:1997,Korotkov:2001,Averin:2005}.
The double dot is operated in the Coulomb blockade regime close
to a charge degeneracy point, where maximally a single
additional electron is allowed to enter and leave the double quantum dot. The Hamiltonian of the full setup is
\begin{equation}
\hatH=\hatH_{DD}+\hatH_{QPC}+\hatH_{DD-QPC}+\hatH_T+\hatH_L+\hatH_R+\hatH_B+\hatH_{DD-B}
\end{equation}
where the various terms are defined in the following.

The Hamiltonian of the double quantum dot is
$\hatH_{DD}=\frac{\varepsilon}{2}\hat{s}_z+T_c\hat{s}_x$, introducing the
pseudo-spin operators $\hat{s}_z\equiv |L\rlangle L|-|R\rlangle
R|$ and $\hat{s}_x\equiv |L\rlangle R|+|R\rlangle L|$.
The tunnel coupling between the two quantum dot levels $|L\rangle$
and $|R\rangle$ is denoted by $T_c$, while $\varepsilon$ is the energy detuning of the two levels. The
pseudo-spin system is tunnel-coupled to left ($L$) and right ($R$)
leads via the tunnel-Hamiltonian
$\hatH_T=\sum_{k_{\alpha},\alpha=L,R}(V_{k_{\alpha}}\hat{c}^{\dagger}_{k_\alpha}
|0\rlangle
\alpha|+\mathrm{h.c.})$, with both leads described as
non-interacting fermions, i.e.,
$\hatH_{\alpha}=\sum_{k_\alpha}\varepsilon_{k_\alpha}\hat{c}^{\dagger}_{k_\alpha}
\hat{c}_{k_\alpha},
\alpha=L,R $. We furthermore include  a dissipative environment consisting of a reservoir of
non-interacting bosons $\hatH_B=\sum_j\hbar\omega_j\hat{a}_j^{\dagger}\hat{a}_j$. The heat bath
 couples to the $\hat{s}_z$ component of
the pseudo-spin via the term $\hatH_{DD-B}=\hat{V}_B\hat{s}_z$ with
$\hat{V}_B= \sum_j \frac{c_j}{2}(\hat{a}_j^{\dagger}+\hat{a}_{j})$, where $c_j$ is the
electron-phonon coupling strength.

The Hamiltonian of the QPC detector is
$\hatH_{QPC}=\sum_{k,\alpha=L,R}\varepsilon_{k \alpha}\hat{d}^{\dagger}_{k \alpha}
\hat{d}_{k \alpha}+\sum_{k,k'}T_0\  \hat{d}^{\dagger}_{k L}
\hat{d}_{k' R}+ h.c.$, where the first term  models the QPC leads and the
second term describes tunneling between them with a real energy-independent tunneling
matrix element
$T_0$. The interaction between the QPC and the
double quantum dot is given by
$\hatH_{QPC-DD}= \sum_{k,k',j=R,L} \delta T_{j}\   \hat{d}^{\dagger}_{k L}
\hat{d}_{k' R}  |j\rlangle
j| + h.c.$ with $\delta T_{j}$ describing the variation of the QPC barrier
opacity due to a localized electron occupying state $|j\rangle$, $j=L,R$. If $\delta
T_{j}\equiv 0$, the current through the QPC at zero temperature is $I^{QPC}=2\pi T_0^2
\mathcal{D}_L \mathcal{D}_R e^2 V/\hbar$ independently of the state of the qubit. Here
$\mathcal{D}_{L/R}$ are the density of states
of the QPC leads and $V$ is the bias across the QPC. If $\delta T_{j}\neq 0$ and the
qubit
is in state $|j \rangle$, the QPC current at zero
temperature is $I^{QPC}_j= 2\pi (T_0+\delta T_j)^2 \mathcal{D}_L
\mathcal{D}_R  e^2 V/\hbar$. For $\delta T_{R}\neq \delta T_L$ the QPC introduces a
decoherence
mechanism of the charge qubit, because it effectively ``measures'' the right and left
states of the qubit.
In this paper we consider the weakly responding limit
where the QPC current is only slightly modified by the charge state of the qubit, such
that
$|I^{QPC}_R-I^{QPC}_L|\ll I^{QPC}$~\cite{Ruskov:2003}.

To describe charge transport through the charge qubit and the QPC
we follow the scheme outlined in Fig.\ \ref{fig:Model} (see also Ref.\ \cite{Gurvitz:1997}). Here, the
variable
$n$ ($m$) corresponds to the number of electrons counted in the right lead of the qubit (QPC).
Instead of writing the equation of motion (EOM) for the
reduced density matrix resolved with respect to the $(n,m)$ components, it is convenient to introduce the
counting fields $\chi$ and $\phi$, corresponding to $n$ (qubit) and  $m$ (QPC), respectively. We then trace
out the leads of the qubit and the QPC, leading to
an EOM for the reduced density matrix
$\hatsigma=(\hatsigma_{00},\hatsigma_{LL},\hatsigma_{RR},
\hatsigma_{LR},\hatsigma_{RL})^T$ of the double dot and the bath of bosons.
The elements
$\hatsigma_{ij}$  are still operators in the Hilbert space of the
boson bath. This approach is valid to all orders in the tunnel
coupling $T_c$ under the assumption of a large bias across the
system and the QPC charge detector \cite{Gurvitz:1997,Gurvitz:1996}.
The
Liouville equation is then
\begin{equation}
\label{eq:EOMMarkov}
\dot{\hatsigma}(t)=\mathcal{L}(\chi,\phi)\hatsigma(t)-i[\hatH_B+\hat{V}_B\hat{s}_z,
\hatsigma(t)],
\end{equation}
where $\mathcal{L}(\chi,\phi)$ describes the reduced dynamics of the charge qubit
and the QPC, while the other contributions are given by the heat bath Hamiltonian $\hatH_B$ and the
interaction term $\hatH_{DD-B}=\hat{V}_B\hat{s}_z$.
The Liouville operator $\mathcal{L}(\chi,\phi)$ for the
combined qubit-detector system is
 \begin{equation}
\label{eq:Matrix}
\!\!\!\!\!\!\!\!\left(
\begin{array}{ccccc}
    -\Gamma_L+D_0 h(\phi) & 0 & \Gamma_R e^{i\chi} & 0 & 0\\
    \Gamma_L  & D_L h(\phi) & 0 & i T_c & -i T_c\\
   0 & 0 & -\Gamma_R+D_R h(\phi) & -i T_c & i T_c\\
   0 & i T_c & -i T_c & -i\varepsilon -\Gamma_d(\phi) & 0\\
   0 & -i T_c & i T_c & 0 & i\varepsilon -\Gamma_d(\phi)
\end{array}
\right)
 \end{equation}
with $h(\phi)=(e^{i\phi}-1)$ and the energy-independent rates
$\Gamma_{\alpha}=2\pi\sum_{k}|V_{k_{\alpha}}|^2
\delta(\epsilon-\varepsilon_{k_\alpha})$, $\alpha=L,R$. These rates describe
charges entering (leaving) the left (right) quantum dot from (to) the left (right) lead.
The number of electrons $n$ that have tunneled to the right lead is increased by tunnel
processes from the right dot with rate $\Gamma_R$, and the counting factor $e^{i\chi}$ consequently
enters the corresponding off-diagonal element of the matrix in Eq.\ (\ref{eq:Matrix}).

The diagonal terms $D_j (e^{i\phi}-1)$ describe counting of tunneling events in the QPC charge detector with
the tunneling rate $D_j=I_j/e$ depending on the state of the qubit, $|j\rangle$, $j=0,L,R$. The state $|0\rangle$ corresponds to the double dot without an additional electron. The generalized dephasing rate $\Gamma_d(\phi)=[\Gamma_R+(\sqrt{D_L}-\sqrt{D_R})^2-
2 \sqrt{D_L D_R}(e^{i\phi}-1)]/2$ represents the decoherence of the
off-diagonal terms  $\hatsigma_{LR}$ and $\hatsigma_{RL}$, taking into account the dephasing induced by the QPC. For $\phi=0$
it yields $[\Gamma_R+(\sqrt{D_L}-\sqrt{D_R})^2]/2$, the dephasing rate
expected from the coupling of the qubit to the right lead connected to the double dot and the nearby QPC
charge detector \cite{Gurvitz:1997}. If $D_L=D_R$, the QPC  does not detect the position of an electron on the double quantum dot.
In that case, the qubit is not loosing its coherence due to the QPC, but only due to the right lead, which contributes with the decoherence rate $\Gamma_R/2$.
The functional dependence on $\chi$ and $\phi$ contains information about
 correlations in the transport statistics of the combined QPC and qubit system. Such correlations will, however, not be
explored in further detail in this work. Here, we limit ourselves to studies of the transport statistics of the qubit.
We note that it is also possible to extend the formalism to cases where the thermal energy $k_B T$ is comparable to the bias $V$
across the QPC \cite{Ruskov:2003}.

We see that the EOM for $\hatsigma(t)$ in Eq.\ (\ref{eq:EOMMarkov}) clearly is a Markovian GME
 defined on a (infinitely) large Hilbert space due to the inclusion of the bosonic heat bath. To
reduce the dimensionality of the problem we trace out the boson degrees of
freedom, and as we shall see this leads to a non-Markovian GME.
We first consider the electronic occupation probabilities
$\rho_{i}=\mathrm{Tr}_B\{\hatsigma_{ii}\}$, $i=0,L,R$, where
$\mathrm{Tr}_B$ is a trace over the bosonic degrees of freedom. The dynamics of the occupation probabilities follow from
Eq.\ (\ref{eq:EOMMarkov})
\begin{eqnarray}
\label{eq:red1}
\dot{\rho}_0(t)&=[-\Gamma_L+D_0 h(\phi)] \rho_0(t)+\Gamma_R e^{i\chi} \rho_R(t),\\
\label{eq:red2}
\dot{\rho}_L(t)&=\Gamma_L \rho_0(t)+D_L h(\phi)\rho_L(t)+i
T_c\mathrm{Tr}_B\{\hatsigma_{LR}(t)-\hatsigma_{RL}(t)\},\\
\label{eq:red3}
\dot{\rho}_R(t)&=[-\Gamma_R +D_R h(\phi)]
\rho_R(t)-i T_c\mathrm{Tr}_B\{\hatsigma_{LR}(t)-\hatsigma_{RL}(t)\}.
\label{eq_rateeq1}
\end{eqnarray}
The effect of the bath enters only via the dynamics of the off-diagonal elements $\hatsigma_{LR}$ and $\hatsigma_{RL}$. The EOM for $\hatsigma_{LR}$  can be written
$\dot{\hatsigma}_{LR}=-\lambda_+(\phi)\hatsigma_{LR}-i( \hatH_B^{(+)}\hatsigma_{LR} +
\hatsigma_{LR}\hatH_B^{(-)})+ iT_c(\hatsigma_{LL}-\hatsigma_{RR})$,
where $\lambda_\pm(\phi)=\pm i\varepsilon+\Gamma_d(\phi)$ and $\hatH_B^{(\pm)}=\hatH_B\pm \hat{V}_B$.
Formally, this equation can be solved as
\begin{eqnarray}
\nonumber \hatsigma_{LR}(t)=&iT_c\int_{0}^t d\tau
e^{-\lambda_+(\phi)(t-\tau)}e^{-i \hatH_B^{(+)}
(t-\tau)}[\hatsigma_{LL}(\tau)-\hatsigma_{RR}(\tau)] e^{i \hatH_B^{(-)}
(t-\tau)}\\
&+e^{-\lambda_+(\phi)t}e^{-i \hatH_B^{(+)}t}\hatsigma_{LR}(0) e^{i \hatH_B^{(-)}
t},
\label{eq_inhomsolution}
\end{eqnarray}
where the term containing the initial condition $\hatsigma_{LR}(0)$ eventually enters
the inhomogeneity \cite{Flindt:2008}. We will only be considering zero-frequency cumulants and can thus safely neglect this term.
We do not show the similar solution for $\hatsigma_{RL}(t)$,  but it is
important to note that $\hatsigma_{RL}(t)$ is
 not simply the complex conjugate of $\hatsigma_{LR}(t)$ due to the counting fields $\chi$ and $\phi$.
Only in the limit $\chi,\phi\rightarrow 0$, the standard relation between
the off-diagonal elements is reestablished.

Substituting the solutions for $\hatsigma_{LR}(t)$ and $\hatsigma_{RL}(t)$ into
Eqs.\ (\ref{eq:red2}) and (\ref{eq:red3}), we can obtain a
closed system of equations by performing a decoupling of the charge degrees of freedom and the boson
bath. Two possible decouplings are considered:
\begin{enumerate}
\item the standard Born factorization, where the system and the bath degrees of freedom are factorized as $\hatsigma_{ii}\simeq
\rho_{i}\otimes\hatsigma_\beta$ with $\hatsigma_\beta\equiv e^{-\beta
H_B}/\Tr_B\{e^{-\beta H_B}\}$.
\item the so-called state dependent Born factorization \cite{Flindt:2004}, where the heat bath is assumed to
equilibrate corresponding to the given charge state, such that $\hatsigma_{LL/RR}\simeq
\rho_{L/R}\otimes\hatsigma_\beta^{(+/-)}$ with $\hatsigma^{(\pm)}_\beta\equiv e^{-\beta
H_B^{(\pm)}}/\Tr_B\{e^{-\beta H_B^{(\pm)}}\}$. This is equivalent to the standard Born approximation, {\em after} the qubit and heat bath have been decoupled via a polaron transformation at $T_c=0$. 
\end{enumerate}
Here, $\beta=1/k_BT$ is the inverse
temperature. These decouplings
are valid when the bath-assisted hopping rates $\Gamma_B^{(\pm)}(z)$ (proportional to
$T_c^2$) are much smaller than $\Gamma_{L/R}$. Additionally, approximation (i) is only
valid when the strength of the electron-phonon coupling is so weak that the state of the qubit
does
 not affect the equilibrium of the heat bath
$\hatsigma_\beta$. We derive an expression for the memory kernel using
assumption (ii). The result corresponding to assumption (i) can easily be obtained via
the
substitution $\hatsigma_\beta^{(\pm)}\to\hatsigma_\beta$.

The memory kernel $W(\chi,\phi,z)$ for our model, with
$\hatrho=(\rho_{0},\rho_{L},\rho_{R})^T$, is given in Laplace space as
\begin{equation}
\!\!\!\!\!\!\!\! \left(
\begin{array}{*3{c c c}}
  -\Gamma_L+D_0 h(\phi) & 0     & \Gamma_R e^{i\chi}\\
  \Gamma_L  & -\Gamma_{B}^{(+)}(z,\phi)+D_L h(\phi)     & \Gamma_{B}^{(-)}(z,\phi)         \\
  0         & \Gamma_{B}^{(+)}(z,\phi)     & -\Gamma_{B}^{(-)}(z,\phi)-\Gamma_R+D_R h(\phi) \\
\end{array}\right)
\label{eq:kernel}
\end{equation}
Most notably, the bath-assisted hopping rates are
$\Gamma_{B}^{(\pm)}(z,\phi)=T_c^2\{\check{g}^{(+)}[z_\pm(\phi)]+\check{g}^{(-)}[
z_\mp(\phi)]\}$ with $z_\pm(\phi)=z-\lambda_\pm(\phi)$. The particular dependence on the QPC
counting field $\phi$ comes about via the dephasing rate $\Gamma_d(\phi)$.
The bath correlation functions $\check{g}^{(\pm)}(z)$ in Laplace space follow from the corresponding expressions in time domain
which can be written as
\begin{equation}
\label{eq:gpm}
g^{(\pm)}(t)=\mathrm{Tr}_B\{ e^{-i \hatH_B^{(+)} t}\sigma_\beta^{(\pm)} e^{i \hatH_B^{(-)}t}\}.
\end{equation}
Upon the substitution  $\hatsigma_\beta^{(\pm)}\to\hatsigma_\beta$, the result for approximation i) is obtained. In that case we have $\Gamma_{B}^{(+)}(z,0)=\Gamma_{B}^{(-)}(z,0)$.

We proceed by calculating the bath correlation functions $g^{(\pm)}(t)$ using assumption i). It is convenient to change representation in Eq.\ (\ref{eq:gpm}), writing
\begin{eqnarray}
g^{(\pm)}(t)&= \mathrm{Tr}_B\{e^{-i \hatH_B^{(\pm)} t}
e^{i \hatH_B t}
\hatsigma_\beta
e^{-i \hatH_B t}
e^{i \hatH_B^{(\mp)} t}\}
\nonumber\\
&= \mathrm{Tr}_B\{\hat{U}_{(\pm)}^\dagger(t)\hatsigma_\beta
\hat{U}_{(\mp)}(t)\},
\label{eq:GKeld}
\end{eqnarray}
where we have introduced the operators
$\hat{U}_{(\pm)}(t)=e^{-i \hatH_B t} e^{i \hatH_B^{(\pm)} t}$, and moreover  used the
fact that $\hatsigma_\beta$ does not evolve with the stationary
Hamiltonian $\hatH_B$. It is easy to demonstrate that the EOM for the
operators $\hat{U}_{(\pm)}(t)$ in this representation is $\partial_t\hat{U}_{(\pm)}(t)=\mp i
\hat{V}_B(t) \hat{U}_{(\pm)}(t)$ with $\hat{V}_B(t)=e^{-i \hatH_B t} \hat{V}_B e^{i \hatH_B t}$. We can thus identify these operators
with the evolution operators in the interaction
picture. The solution of the
time dependent differential equations is ${\hat{U}}_{(\pm)}(t)=\hat{T}[ e^{\mp i\int_0^t
d\tau V_B(\tau)}]$
with $\hat{T}[\cdot]$ being the time-ordering operator. Equation (\ref{eq:GKeld}) can then be written
\begin{eqnarray}
g^{(\pm)}(t)&=\mathrm{Tr}_B\{\tilde{T}[e^{\mp i\int_0^t d\tau V_B(\tau)}]\ \rho_\beta\
\hat{T}[e^{\mp i\int_0^t d\tau \hat{V}_B(\tau)}]\}\nonumber\\
&\equiv\Tr_B\{T_K[e^{\mp i\int_0^t
dt' \tau_3
V_B(t')}]\},
\end{eqnarray}
where we have introduced the time-antiordering operator $\tilde{T}[\cdot]$ and the Keldysh contour ordering operator $T_K[\cdot]$.
The second equality expresses
the bath correlation function in terms of a  Keldysh propagator with a branch
dependent interaction potential $\tau_3
V_B(t')$ with $\tau_3=+(-)$ for the forward (backward) Keldysh branch. This
expression can be evaluated using perturbation theory in the electron-phonon couplings $c_j$ and we
obtain $g^\pm(t)=1-\int_0^{\infty}d\omega
J(\omega)\{[1-\cos(\omega
 t)]\coth(\beta\omega/2)\}/\omega^2+O(|c_j|^4)$,
where $J(\omega)=\sum_j |c_j|^2\delta(\omega-\omega_j)$ is the spectral density of
the heat bath.  Below, we consider the case of Ohmic dissipation $J(\omega)= 2 \alpha\omega
e^{-\omega/\omega_c}$. We note that $g^{(+)}(t)=g^{(-)}(t)\equiv g(t)$.
The Laplace
transform of $g(t)$, to leading order in
$1/\beta\omega_c$, reads
\begin{equation}
\check{g}(z)=\frac{1}{z}
\left\{1-2\alpha
    \left[\ln\left(\frac{\beta z}{2\pi}\right)
          -\Psi\left(\frac{\beta z}{2\pi}\right)
          -\frac{\pi}{\beta z}
          -\mathcal{D}\left(\frac{z}{\omega_c}\right)
    \right]
\right\},
\label{eq:GZ}
\end{equation}
where $\Psi(x)$ is the digamma
function and $\mathcal{D}(x)=\mathrm{ci}(x)\cos(x)+\sin(x)[\mathrm{si}(x)-\pi/2]$ with
the sine and cosine integrals defined as
$\mathrm{si}(x)=\int_0^x dt \sin(t)/t$ and
$\mathrm{ci}(x)=-\int_x^\infty dt \cos(t)/t$, respectively.

For decoupling ii) we can calculate the
bath correlation functions exactly to all orders in $\alpha$ using standard many-body
techniques~\cite{Flindt:2008a}. The results for the bath correlations functions are
$g^{(\pm)}(t)=e^{-W(\mp t)}$, where
\begin{equation}
\label{eq:Wdiss}
 \qquad W(t)=\int_0^{\infty}d\omega
\frac{J(\omega)}{\omega^2}\left\{[1-\cos(\omega
 t)]\coth\left(\frac{\beta\omega}{2}\right)+i\sin(\omega
t)\right\}.
\end{equation}
In the weak coupling limit, we
can expand this expression to first order in $\alpha$. The
bath correlation functions $\check{g}^{(\pm)}(z)$ for an Ohmic spectral density,
to leading order in
$1/\beta\omega_c$, are $\check{g}^{(\pm)}(z)=\check{g}(z)\pm i 2 \alpha
F(z/\omega_c)/z$ where $\check{g}(z)$ is given in Eq.\ (\ref{eq:GZ}) and
$F(x)=\mathrm{ci}(x)\sin(x)+\cos(x)[\pi/2-\mathrm{si}(x)]$. We see that the difference between the bath correlation
functions obtained within the two decoupling schemes is proportional to the imaginary
of part Eq.\ (\ref{eq:Wdiss}). This difference is the main reason that
decoupling i) does not lead to any asymmetry of the cumulants between the
emission and absorption sides, as we shall see. We note that the bath correlation function is
$\check{g}(z)=1/z$ for $\alpha=0$. This
corresponds to the coherent regime of the qubit  and the $z$-dependence of
$\Gamma_{B}^{(\pm)}(z,\phi)$  describes the ``effective'' memory due to the off-diagonal elements that have been traced out. Furthermore, for $z\to 0$ the rates become
$\Gamma_{B}^{(\pm)}(0,\phi)=4T_c^2\Gamma_d(\phi)/[\Gamma_d(\phi)^2+4\varepsilon^2]$,
and for $\phi=0$ we obtain the standard expressions for incoherent tunneling rates
in a double dot \cite{Kiesslich:2007}.

We now consider the current cumulants of the charge qubit for $\phi=0$, where the effects of the
QPC is captured by the dephasing rate $\Gamma_d$. Evaluating Eqs.\ (\ref{eq:I}) and (\ref{eq:III}) using the kernel in Eq.\ (\ref{eq:kernel})
we find the analytic expressions for the cumulants. The expressions for the current and noise with $\alpha=0$ coincide with known results for the coherent case~\cite{Gurvitz:1996,Stoof:1996}, however, with the total dephasing rate being $\Gamma_d=[\Gamma_R+(\sqrt{D_L}-\sqrt{D_R})^2]/2$ rather than just $\Gamma_R/2$. Considering only the Markovian contributions $c^{(k,0)}$, we obtain the well-known analytic
expressions for sequential tunneling \cite{Kiesslich:2007}. We next compare different regimes with various strengths of the QPC dephasing rate $\Gamma_d$ and
different electron-phonon couplings $\alpha$.
In Fig.\ \ref{fig:Model}b, we show the mean current as function of the level detuning $\varepsilon$. The
solid black line corresponds to the coherent regime, where the peak is symmetric around
$\varepsilon=0$ and the width of the peak is proportional to $\Gamma_R/2$. The dotted
line shows the effect of the dephasing due to the QPC charge detector, without any contributions from
the boson bath ($\alpha=0$). For $D_R \neq D_L$, the
peak width is given by the total dephasing rate, i.e., the sum of the
intrinsic contribution  $\Gamma_R/2$ and the contribution from the QPC
$(\sqrt{D_L}-\sqrt{D_R})^2/2$. The width is thus larger than in the coherent case.
The peak, however, remains symmetric.
The gray curves show the mean current, when the double dot is coupled to the
heat bath at finite temperature, but with the QPC charge detector not detecting the position of an electron on the double dot ($D_R=D_L$). The two
gray lines
correspond to the two different decouplings, decoupling i) with
light gray and decoupling ii) with dark gray. We note that the curve corresponding to decoupling ii) is asymmetric around
$\varepsilon=0$. Finally, the dot-dashed curve corresponds to the mean current with contributions to the dephasing  from both the QPC and the heat bath.

\begin{figure}
\begin{center}
\includegraphics[width=.9\linewidth]{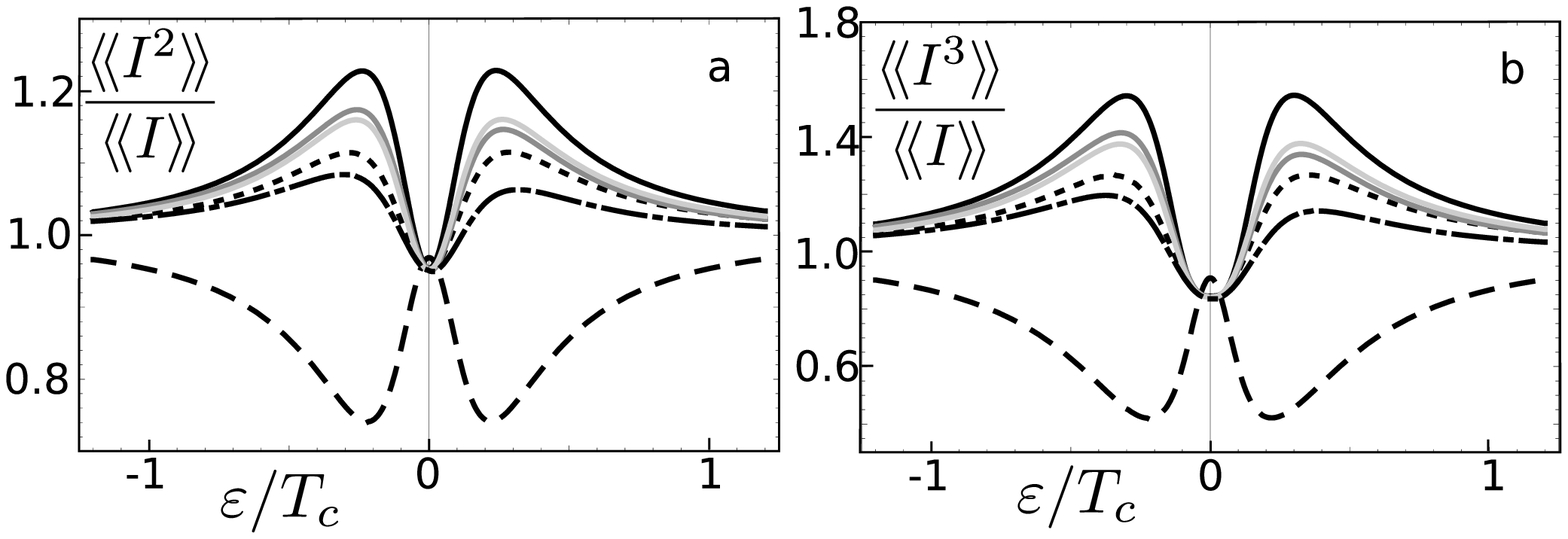}
\caption{a) Fano factor $\llangle I^2 \rrangle/\llangle I\rrangle$ as a function of
the level detuning $\varepsilon/T_c$. b) Normalized skewness $\llangle I^3 \rrangle/
\llangle I\rrangle$ vs $\varepsilon/T_c$. Parameters corresponding to the different line styles are given in the caption of Fig.\ \ref{fig:Model}.}
\label{fig:NoiseSkew}
\end{center}
\end{figure}

While the mean current studied so far only reveals little information about the dephasing mechanisms, we expect more information to be contained in the higher-order cumulants.
In Fig.\ \ref{fig:NoiseSkew} we show the Fano factor $\llangle I^2 \rrangle/\llangle I\rrangle$
and the normalized skewness $\llangle I^3 \rrangle/\llangle I\rrangle$ as functions of
the level detuning $\varepsilon$. Line styles and parameters are the same as in
Fig. \ref{fig:Model}b. Generally, we observe that the two cumulants to a higher degree than the mean current discriminate between various regimes. In particular, the coherent regime has strongly super-Poissonian behavior with both of the normalized cumulants reaching values larger than the Poissonian limit of $1$. In contrast, for the sequential tunneling regime, given by the Markovian contributions to the
cumulants, only sub-Poissonian behavior is observed \cite{Kiesslich:2007}.

If a dephasing mechanism is introduced, either due to the QPC charge detector or the heat bath,
the super-Poissonian behavior is gradually reduced towards the sequential tunneling regime with increasing dephasing rate. The dephasing rate $\Gamma_d$ due to the QPC can be modified via the coupling between the charge qubit and the QPC. On the other hand, the dephasing induced by the heat bath changes with the bath
temperature, which can also induce the transition between
coherent and incoherent transport. Such a transition was recently observed in
 shot noise measurements of transport through a double dot at low temperatures \cite{Kiesslich:2007}. Comparing the scales of the vertical axes of the second and third cumulants in Fig.~\ref{fig:NoiseSkew}, we conjecture that such a transition may be better visible in the third cumulant and that even higher order cumulants in general could be more sensitive to such a transition.

We now study, how the symmetry of the cumulants changes in the various regimes. Without coupling to the QPC or the heat bath the cumulant are symmetric around $\varepsilon=0$. In case the qubit is coupled only to the QPC, this symmetry remains intact. However,
with non-zero coupling to the heat bath, asymmetry around $\varepsilon=0$ is observed (see dark gray and
dot-dashed lines). The asymmetry occurs due to the
asymmetry in emission and absorption of bosons at low temperatures. Phonon absorption and emission dominate for $\varepsilon<0$ and $\varepsilon>0$, respectively. We note
that the curves obtained using decoupling i) do not capture this essential physics, not even to first order in $\alpha$. Indeed, in all the regimes for
which the cumulants are symmetric, the rates have the  property
$\Gamma^{(+)}(z)=\Gamma^{(-)}(z)$. This particular property is not fulfilled in decoupling ii) due to the imaginary part of
Eq.\ (\ref{eq:Wdiss}), and we believe that decoupling ii) should be correct to any order in $\alpha$. What is then the essential difference between the dephasing
induced by the QPC charge detector and the bath-induced dephasing in the model we are
considering here? The QPC charge detector induces dephasing with little influence on the
dynamics of the system and does not destroy the symmetry in the coherent regime.
In contrast, the bath-induced dephasing is generated with emission and absorption of bosons, and
the charge qubit is influenced by the intrinsic asymmetry of this process. We conclude the
analysis of this model for now and postpone a further analysis to future research aimed at an improved understanding of these dephasing mechanisms.

\section{Outlook and open questions}
\label{sec:Open}

We have shown how it is now possible to calculate cumulants, also of high orders,
for a wide class of counting problems. In this work we have applied the method to investigate how the dephasing mechanism of a QPC charge detector differs from that of an external heat bath. The method is, however, applicable to a large class of non-Markovian counting problems, also from outside the field of physics. The general approach outlined here is also suitable for calculations of the finite frequency noise spectrum \cite{Flindt:2008}. It is relevant to ask if it is possible to extend this approach to frequency dependent cumulants of higher orders, similar to the results already obtained for Markovian systems \cite{Emary:2007}? Another interesting issue concerns systems with very long memory time (for example, Levy-flight processes). Such cases would require us to reconsider some of the assumptions underlying the theory presented here.

We would like to thank R.~Aguado, T.~Brandes, A.-P.\ Jauho, S.~Kohler,
K.~Neto\v{c}n\'{y}, and M.\ Sassetti for fruitful discussions and suggestions. The
work was supported by INFM-CNR via ``Seed'' project, the Villum Kann
Rasmussen Foundation, the grant 202/07/J051 of the Czech Science Foundation, and research plan MSM
0021620834 financed by the Ministry of Education of the Czech Republic.

\end{document}